\documentstyle[epsfig]{aipproc}
\newcommand{\gray}{$\gamma$-ray}
\newcommand{\grays}{$\gamma$-rays}
\newcommand{\etal}{{\it et al.}}

\newcommand{\amsg}{AMS/$\gamma$}

\begin{document}

\title{The Capabilities of the Alpha Magnetic Spectrometer as GeV $\gamma$-Rays Detector}


\author{R. Battiston\footnote{Invited talk at the Workshop
 on "GeV-TeV Gamma-Ray Astrophysics", 13-16 August 1999, Snowbird, Utah.}}
\address{Sezione INFN and Dipartimento di Fisica dell' Universit\`{a},
Perugia, Italia 06100}
%
\maketitle

\begin{abstract}
The modeled performance of the Alpha Magnetic Spectrometer (AMS) as a high-energy
(0.3 to 100 GeV) gamma-ray detector is described, and 
its gamma-ray astrophysics objectives are discussed.
\end{abstract}


\section*{Introduction}
\label{intro.sec}

Our knowledge of the \gray\ sky has increased dramatically during this
last decade, due principally to the \gray\ instruments on board the
Compton Gamma Ray Observatory (CGRO)\cite{kur97}: EGRET,
a spark chamber plus calorimeter instrument with \gray\ sensitivity in the
energy interval 30 MeV to 30 GeV; COMPTEL, 

\begin{figure}[ht]
\begin{center}
\mbox{\epsfig{file=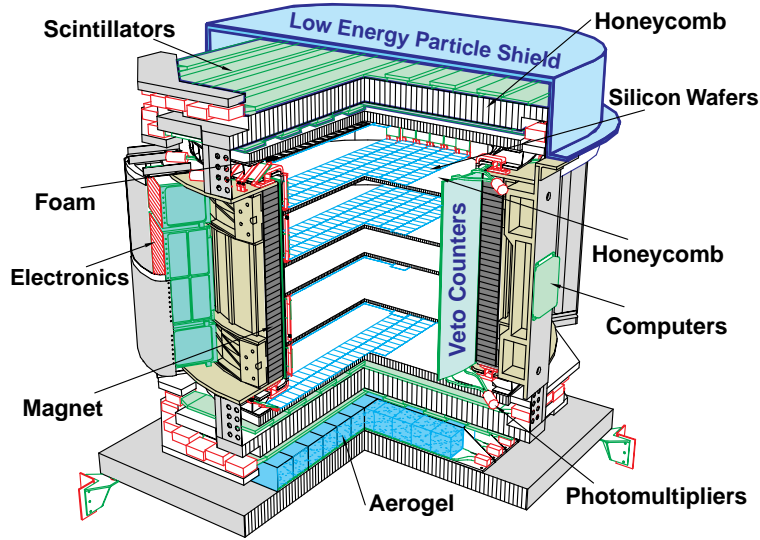,width=10cm}}
\caption{\small A cross section of the baseline AMS instrument.}
\label{ams-nim.fig}
\end{center}
\end{figure}

a Compton telescope in the interval 0.1 to 30 MeV; BATSE, an
omnidirectional x-ray and soft \gray\ ``burst'' detector consisting of 
large NaI scintillators sensitive to 30 keV to 2 MeV photons (with smaller
spectroscopic NaI crystals for measurements up to 110 MeV); and 
OSSE, consisting of Nai-CsI phoswiches detecting photons of
0.1 to 10 MeV.  Their observations
have revolutionized our understanding of such extragalactic phenomena
as blazars and gamma ray bursts (GRBs), as well those within our
own Galaxy, such as pulsars.

Until recently, observations with CGRO and other space-based
and ground-based telescopes have provided coverage of these
and other sources up to the limiting sensitive energy of EGRET, approximately
30 GeV.  From there, a gap in our knowledge of the \gray\ sky spectrum
has existed up to 200-300 GeV, where ground-based \gray\ shower detectors
presently have their energy thresholds.  It is possible that within this
gap there are novel features in the \gray\ sky, such as a gamma-ray line
or continuum emission from postulated neutralino annihilation at the 
center of the Galaxy \cite{jun96},\cite{ull97},\cite{ber97a}.  Future instruments with sensivity in this
unexplored region, such as  AGILE \cite{agi98}, recently approved by
the Italian Space Agency (ASI),  or  GLAST \cite{eng97} being proposed to 
NASA, may uncover exciting new   phenomena.

\begin{figure}[ht]
\begin{center}
\mbox{\epsfig{file=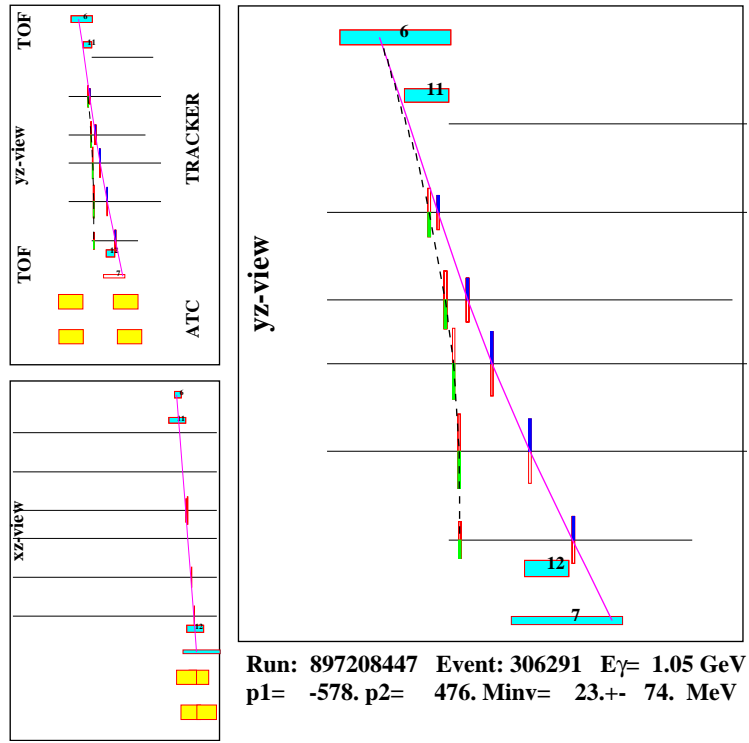,width=10cm}}
\caption{\small A $1 GeV$ \gray\ conversion detected during the STS91 AMS  flight.}
\label{gammaevent.fig}
\end{center}
\end{figure}

At the present time, however, our view of the \gray\ sky has diminished.
With the effective turnoff of the EGRET \gray\ instrument on the
CGRO due to the nearly
complete consumption of its spark chamber gas, there is
no operating instrument capable of observing the \gray\ sky in
the energy interval $\sim 10^{-1}$ to $\sim 10^{2}$ GeV.
Ground-based \gray\ detectors, based on the atmospheric Cerenkov
technique (ACT)\cite{hil90}, turn on at current energy thresholds 
of $\sim$200-500 GeV.  Within
the next several years
energy thresholds for some ACT observatories are expected to go to as low as
20 to 50 GeV, but lower
thresholds than these are unlikely to be achieved due to the
sizable effect of Earth's magnetic field on the \gray -induced
air showers, and the lower Cerenkov photon yield which must be
detected against the night sky background.  For \gray\ energies much lower
than $10^{2}$ GeV, then, spaced-based detectors are required.  
This observational gap for the energy window $\sim 10^{-1}$ to $10^{2}$
GeV will continue to exist for the next several years.
Eventually, this gap will be eliminated
with the launch of a next-generation \gray\ satellite, such as GLAST \cite{eng97},
but such a mission is unlikely to occur before the year 2005.  Also the
AGILE satellite \cite{agi98}, to be launched in 
2002, will have  a limited sensitivity above
50 GeV.

In this paper we describe how the Alpha Magnetic Spectrometer (AMS)
can largely fill this gap by acting as a \gray\ detector with sensitivity in the
energy interval of 0.3 to 100 GeV during its 
three-year mission on board the
International Space Station Alpha (ISSA) from 2003 to 2006.
AMS, described in detail elsewhere \cite{ahl94}, has as primary mission
the search for cosmic ray antinuclei as well as the
search for dark matter studying  anomalies in CR spectra
and composition (e.g. $e^+,\bar{p}$). 

\begin{figure}[ht]
\begin{center}
\mbox{\epsfig{file=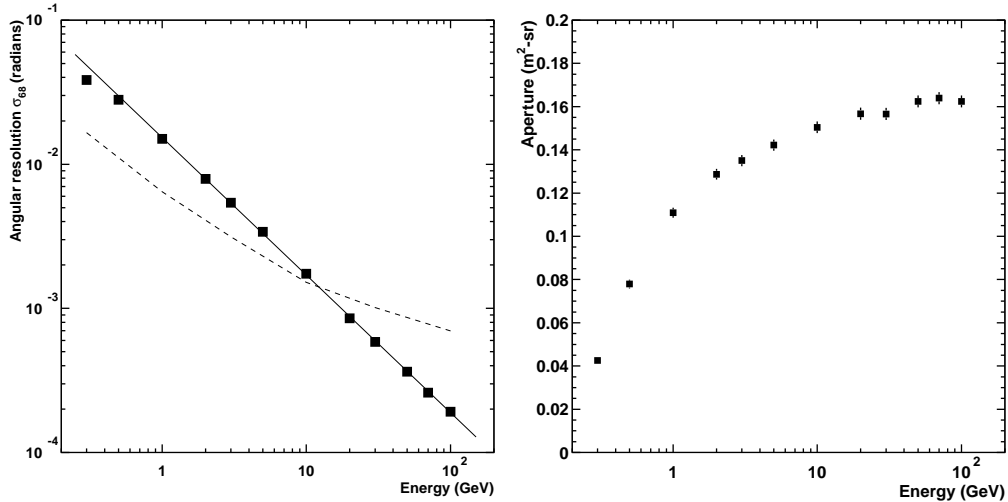,width=14cm}}
\caption{\small (a) Angular resolution of \amsg\ 
 (filled squares) and   of GLAST  (dashed line)
 as a function of primary \gray\  energy  in the interval 0.3 to 100 GeV;(b)
\amsg\ aperture as a function of \gray\ energy.}
\label{angres.fig}
\end{center}
\end{figure}

\begin{figure}[ht]
\begin{center}
\mbox{\epsfig{file=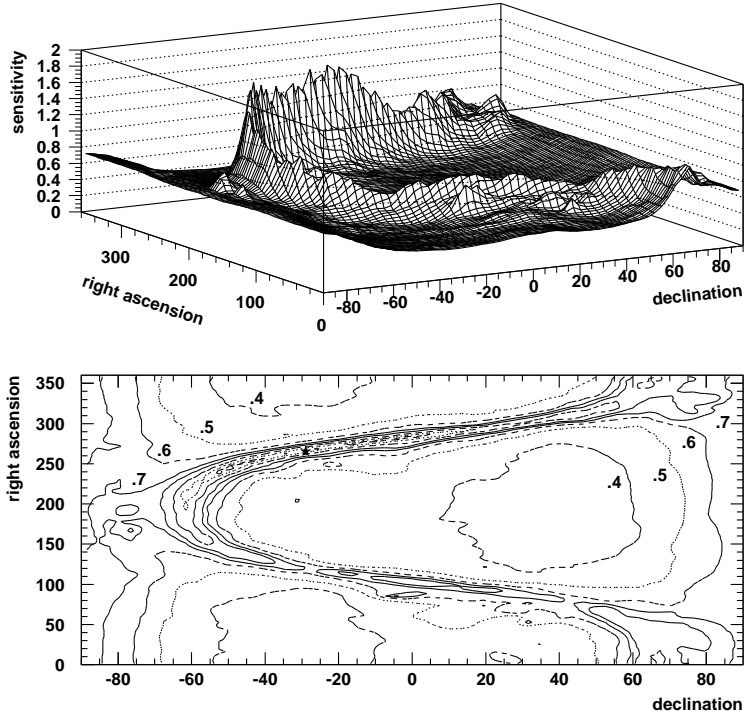,width=11cm}}
\caption{\small A 3-D plot and 2-D contour plot of \amsg 's point source
sensitivity $n_{0}$ versus celestial coordinates.  Recall that $n_{0}$ is
the minimum amplitude of the source's differential flux at 1 GeV required for
a $5\sigma$ significance detection.  The units for $n_{0}$ in the 3-D plot are
$10^{-8}$ cm$^{-2}\cdot$s$^{-1}\cdot$GeV$^{-1}$.  Large photon fluxes from the diffuse
galactic background
are responsible for the deterioration of sensitivity near the Galactic plane.}
\label{sens.fig}
\end{center}
\end{figure}

 There are at least  two ways  this modification could be done  without significantly affecting
the experiment sensitivity to antimatter: (a) adding a light (e.g. $0.3\ X_0$)
converter at the entrance of the magnetic spectrometer, either passive
(e.g., a high-Z thin plate) or active (e.g. a multilayer tracking detector), and/or
(b) implementing an high granularity imaging shower detector at the bottom
of the experiment. 
In this paper we  present a study which has been perfomed on option (a). We
will show that in this  option, AMS can also detect \grays\ with
performance characteristics similar to EGRET in
the energy region of 0.3 to 20 GeV, and with significantly enhanced
capabilities between 20 and $\sim$ 100 GeV, a region which is not well
explored.  We refer to this modified instrument as ``\amsg''.
We show that \amsg\ can continue the valuable work of EGRET by providing
continued monitoring of extragalactic and galactic \gray\ sources and by
participating in multiwavelength observational campaigns.  In addition,
\amsg\ will have unprecedented sensitivity to the \gray\ sky between the
energies of 20 to $\sim 100$ GeV (albeit at a level somewhat lower than required
for detection of known sources with power law spectra), so that
\amsg\ might provide us with unexpected discoveries in this region.
 The next section gives a brief description of the baseline AMS instrument and its experience
during the first precursor flight.
Then we describe the performance characteristics of \amsg\ in option (a) as determined from
Monte Carlo analyses and we perform a comparison  to EGRET. 
For a more detailed description of the  results of these studies we refer to \cite{gammapaper}.

\section*{The Alpha Magnetic Spectrometer on the STS91 flight}
\label{baseline.sec}

The Alpha Magnetic Spectrometer has been built by a large international
collaboration of high energy physics institutions from the U.S., Italy,
China, Finland, France, Germany, Taiwan, Russia, and Switzerland.
It recently had a successful test flight on the Space Shuttle mission STS-91 in June 1998
\cite{bat99},
when it was carried in the cargo bay and observed for several days in both
the zenith and nadir directions, the latter for measurement of albedo
cosmic ray backgrounds. The design of the baseline AMS instrument flown during the STS-91 mission 
is shown in Figure \ref{ams-nim.fig}.  

The magnet spectrometer consists of a permanent ring dipole magnet made of
very high grade Nd-B-Fe rare earth material whose magnetic energy product
and residual induction are respectively $(BH)_{\rm max}>50\times 10^{6}$
G-Oe and 14,500 G, yielding an analyzing power of $\left<BL^{2}\right>
=0.14$ T-m$^{2}$ with less than 2 tonnes of magnet mass.  
Four high precision silicon strip detector tracking planes are located
within the magnetic volume, with a fifth and sixth plane located
just above and below the magnet \cite{bat95}, \cite{bur98}.
At low rigidities (below 8 GV) the  resolution  of the spectrometer is dominated by multiple 
scattering  ($\Delta p/p\sim 7\% $), while the maximum detectable rigidity ($\Delta p/p\sim 100\%$) 
is about $500$ GV. 
In addition to measuring particle rigidity, the silicon planes will provide six
independent measurements of $dE/dx$ for charge determination. 
Four time-of-flight (ToF) scintillator planes (two above and two below the
magnet volume) 
measure the particle  velocity with a resolution of
120 ps over a distance of 1.4 m.  
The ToF scintillators also  measure $dE/dx$, allowing a
multiple determination of the absolute value of the particle charge.

A solid state Cerenkov detector below the magnet provides an
independent velocity measurement, useful to separate electrons and positrons from 
the hadronic  CR components  (protons, helium...) and residual background (pions).
In addition, a scintillator anticounter system
is located within the inner magnet wall, extending to the ToF scintillators.

The performance of this instrument as a charged  cosmic ray detector, and its
sensitivity to an antinucleus cosmic ray component, is discussed
elsewhere\cite{ahl94}.

 During the STS-91 AMS has collected about 50 million of 
single CR events above $\sim 100 MeV/n$ kinetic energy,
 improving the existing cosmic antimatter  limits \cite{amsaHe} and measuring 
in details the structure of the CR flux at 400 km of height and over most of the 
surface of the earth \cite{amsFlux}. 

A few tens of clean  two tracks  events compatible with GeV-range
\gray\ converting  in to an $e^+e^-$ pair on  the top layers ($\sim 5\%\ Xo$)
of the spectrometer have also been  observed  during the precursor flight (see Figure \ref{gammaevent.fig}).
This is, to our knowledge, the first time high energy $(E_{\gamma}> 1\ GeV))$ \grays\ 
are observed in a magnetic spectrometer in space:
thanks to the dipolar magnetic field  the $e^+e^-$ pair opens  up in the bending plane view but it  remains very 
collimated in the non-bending view as it is clearly shown  in the Figure. In addition to the particle 
identification capabilites of the AMS detector, these   distinctive topological features  
of a \gray\ conversion will be important  in  rejecting the  background induced by the  $O(10^5)$ times larger  
flux of charged CR. 

 AMS  is scheduled to be secured to an
external payload attachment point on the International Space Station Alpha
in may 2003, where it will remain as a zenith-pointing instrument for 
3 years of measurement time. We should also note that by the time AMS is attached to the International
Space Station it may have undergone significant changes from the baseline
design considered here.  In particular, the permanent magnet may be replaced
by a superconducting magnet, which would considerably improve
AMS's \gray\ detection performance at high energy.  Since the
detector is still undergoing significant design changes from the baseline
instrument flown on the Space Shuttle, we have chosen to fix on that design
which currently exists as integrated hardware.  The addition of other
components discussed in the proposal and currently under developement,
 such as a transition radiation detector located at the entrance of the magnet 
and a solid state \v{C}erenkov radiator  followed by a segmented calorimeter located at the bottom of the instrument,
 will   improve the  particle identification capability of  AMS as well as its  performance as 
 \gray\ detector.

\section*{The Performance of AMS/$\gamma$}
\label{performance.sec}

In option (a), the conversion of AMS to \amsg\ requires the addition of 
two hardware components: (1) A  passive  or active   converter  medium, located
at the entrance of the spectrometer which
converts \grays\ into electron-positron pairs.   (2) A stellar attitude
sensor gives the angular orientation of AMS with respect to the celestial
sphere to an accuracy of better than $1.5\times 10^{-4}$ radians.

The determination of the converter thickness is based on an optimization between the
probability  of \gray\ convertion, the amount of bremstrahlung losses, which  limit the energy resolution,   and the 
amount of multiple scattering, which limits the angular resolution. 
In addition this  modification should not degrade the sensitivity of AMS to the
search of antimatter and its particle identification capability. 
  Following the  results of our study we choose the value $x=0.3X_{0}$, 
for which point source sensitivity is still optimal,
the energy resolution is acceptable, and nuclear interaction losses are
negligible. For the scope of our study we assumed a $x=0.3X_{0}$ tungsten plate converter
located  before the first ToF layer.

A full-instrument GEANT Monte Carlo code was run to determine the 
performance of \amsg. Gamma rays with fixed energies, ranging 
from 0.3 to 100 GeV, were
thrown isotropically at the detector over an opening angle of 
$50^{\circ}$. 
All the
physical processes for electrons and \grays\ were ``on'' in the GEANT
code.  Bremsstrahlung photons of energies $<20$ MeV were not followed,
although all bremsstrahlung energy losses were included.  Cuts simulating the trigger conditions and  the pattern
recognition algorithm were applied on the converted events. 

Primary \gray\ energy and incidence direction were determined by adding the fitted
momenta vectors of all secondaries evaluated at the converter plate to obtain
the primary momentum vector.  

Particular care has been given to the study of  the instrumental background
 generated by CR simulating a \gray\ conversion. As reference for  this study we used
 the Extragalactic Gamma Ray Background (EGRB) at High Energies. 
 Any measurement of the EGRB requires that instrumental background events
be kept to a lower rate than the EGRB flux.  To estimate the effects
of background due to collisions of cosmic ray electrons, positrons, and
protons with the AMS instrument, we divided the \gray\ spectrum into
four bins per decade of energy (from 0.5 to 100 GeV) and required that
we investigate backgrounds down to a level of 20\% of the EGRB rate
in {\it each} energy bin.  For example, in the 25-40 GeV bin it will
take \amsg\ $2.0\times 10^{5}$ seconds to obtain 5 EGRB \grays.  In our
Monte Carlo analysis, therefore, we threw $2.0\times 10^{5}$ seconds' worth of
cosmic ray electron, positron, and proton flux at the instrument to
generate a false \gray\ background.  These cosmic rays were thrown isotropically
over a zenith angle range of $0^{\circ}<\theta <110^{\circ}$, where the
largest zenith angle corresponds to the location of the Earth limb at
an orbital altitude of 400 km.  The energies were distributed according
to known electron\cite{mul87},\cite{gol94}, positron\cite{bar97}, and proton\cite{men97}
 energy spectra.

  Using  the same quality cuts  to eliminate the electron-induced  and proton-induced background events,
   we had a total of 2 electron-induced and  5 proton induced  events; all events had reconstructed 
   energies of $<1$ GeV, leaving no background events in the interesting high energy region.

\subsection*{Comparison of \amsg\ with EGRET}

Since the basic parameters of \amsg\ are close to those  of EGRET  
in the following we  present a comparison  among the  two detectors. 
The converter thickness
largely dominates all multiple scattering effects,
so that the angular and energy resolution of reconstructed primary photons
will be completely dominated by multiple Coulomb scattering (MCS) and bremsstrahlung energy
losses of the electrons within the converter plate.  This is confirmed by the 
full MC simulation which gives an energy dependence on the  angular resolution:  
\begin{equation}
\sigma_{68}^{\rm AMS}(E)=0.88^{\circ}\left(\frac{E}{1 {\rm \ GeV}}\right)^{-0.956},
\label{eq6}
\end{equation}
which is to be compared to EGRET's angular resolution \cite{tho93} of
\begin{equation}
\sigma_{68}^{\rm EGRET}(E)=1.71^{\circ}\left(\frac{E}{1 {\rm \ GeV}}\right)^{-0.534}.
\label{eq7}
\end{equation}
In Figure \ref{angres.fig}a
the expected \amsg\ angular resolution is compared to the corresponding figure for GLAST\cite{blo96}, where 
we note, as it should be expected, that at sufficiently high energy a precise pair spectrometer does  
eventually give a better angular  resolution than a fine grain  imaging calorimeter. 

An integration of effective area $A(E,\theta)$ over solid angle gives 
instrument aperture, shown in Figure \ref{angres.fig}b.
Below a \gray\ energy of 0.3 GeV, the converted electrons begin to have too small a radius
of curvature to escape from the magnet volume, and detection efficiency plummets.
Above 100 GeV 
the converted electron and positron often do not spatially diverge beyond the
two-hit resolution distance of the silicon trackers, causing significant deterioration
in \gray\ energy resolution.  These considerations
then define the limits of the energy window for \amsg.

One cannot directly compare the point source sensitivity of \amsg\ 
to that of EGRET, since AMS is not a pointable instrument.
EGRET achieves a flux sensitivity of $I_{\rm min}(>0.1 {\rm \ GeV})
\approx 10^{-7} {\rm cm}^{-2}{\rm s}^{-1}$ with a 2-week viewing period.
Since it took EGRET one year to map the full sky, where each sky segment
was viewed for roughly 2 weeks, we can compare EGRET's sensitivity with that 
achieved by \amsg\ after one year of operation.
By assuming a source differential spectrum of $E^{-2}$, we can convert from
EGRET's definition of sensitivity (in terms of integral flux above 0.1 GeV)
to that of \amsg\ (a differential flux above 1 GeV).  In our units the
EGRET $5\sigma$ flux sensitivity is $1\times 10^{-8}$.
Over most of the sky, \amsg 's 
mean sensitivity $\left<n_{0}\right>$ is estimated to
be about a factor of 2 lower than that of EGRET (Figure \ref{sens.fig}).

 The table below summarizes the performance characteristics of AMS as a \gray\ 
detector.  In particular, by its comparison to EGRET, one sees that
the two instruments perform similarly in many respects, one major difference being
the energy windows: AMS's energy window is shifted up by roughly one order of magnitude from
that of EGRET, thereby providing an improved view of the sky in the region
$E_{\gamma}\sim 100$ GeV.\\

\begin{tabular}{|l|c|c|} \hline
 & AMS & EGRET \\ \hline \hline
technique & magn.spectrometer & spark ch.$+$calorimeter \\ \hline
energy window (GeV) & 0.3 to 100. & 0.03 to 30. \\ \hline
peak effective area (cm$^{2}$) & 1300 & 1500 \\ \hline
angular resolution & $0.77^{\circ}(E/1 GeV)^{-0.96}$ & $1.71^{\circ}(E/1 GeV)^{-0.534}$ \\ \hline
half-area zenith angle & $\sim 30^{\circ}$ & $\sim 20^{\circ}$ \\ \hline
total viewing time (yr) & $\sim$3 & $\sim$2 \\ \hline
attitude capability & fixed & movable \\ \hline
flux sensitivity & $\sim0.5\times10^{-8}$ & $\sim1.0\times10^{-8}$ \\
(ph/cm$^{2}$-s-GeV at 1 GeV) & & \\ \hline
\end{tabular}
\vspace{0.5cm}
\begin{figure}[ht]
\begin{center}
\mbox{\epsfig{file=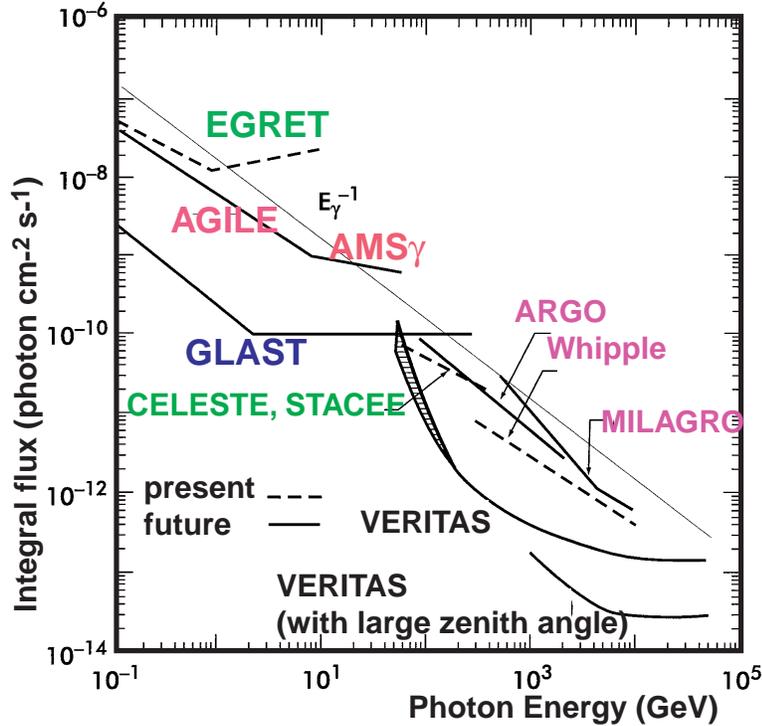,width=12cm}}
\caption{\small Sensitivity of present and future \gray\ detectors.}
\label{sensitivity.fig}
\end{center}
\end{figure}

Note that the point source sensitivity
is nearly the same for both EGRET and AMS,
with that of AMS being somewhat (factor $\sim$2) lower.
This is because EGRET and AMS have 
similar effective detection areas and angular resolutions.
The fact that AMS's \gray\ energy threshold is an order of magnitude
larger than EGRET's, thereby implying an integrated point source flux about an order of
magnitude lower for AMS than for EGRET, is compensated for by the fact that AMS's angular
aperture (half-angle area of $30^{\circ}$) is larger than EGRET's ($\sim 20^{\circ}$),
and by the fact that AMS spends 100\% of its time pointing to the sky
(being attached to a gravity-gradient stabilized Space Station), while EGRET typically points
one-third of the time to Earth. 
AMS lacks the low-energy end of EGRET's range due to the curvature
of the electron-positron pair in AMS's magnetic field, which limits the detectable  \gray\ 
energy to $\geq 300$ MeV.
On the other hand, EGRET lacks the high-energy end of AMS's range due to the effect of
electromagnetic backsplash in EGRET's NaI calorimeter which vetos most events above 30 GeV.

Since  \amsg\ and EGRET have very similar \gray\ detection
capabilities, \amsg\ will be able to continue and extend the investigation of
galactic and extragalactic \gray\ sources initiated by EGRET.  One main difference
between the two detectors, that of significant aperture above 30 GeV for \amsg,
may lead to the observation of new phenomena in this relatively uncharted region.

Figure \ref{sensitivity.fig}  shows a comparison between the sensitivity of present and future high
energy \gray\ detectors. 

\section*{Conclusions}
\label{conclusions.sec}
We have shown that with minor modifications the Alpha Magnetic Spectrometer can
become a powerful \gray\ detector as well, with overall performance characteristics
being comparable, if not superior, to those of EGRET.  With \gray\ energy
resolution extending past 100 GeV, and with an aperture that is nearly 
flat above $\sim 3$ GeV, \amsg\ can address a number of outstanding issues in 
\gray\ astrophysics that relate to the relatively unexplored region of
$E_{\gamma}=20-200$ GeV.  For one, \amsg\ will likely confirm or refute
the hypothesis that unresolved blazars are responsible for the bulk of the
extragalactic \gray\ background; \amsg\ will also extend the spectrum of the
diffuse galactic background to above 100 GeV, helping to resolve current
difficulties in interpreting the EGRET diffuse galactic background 
measurement \cite{poh98}.

\amsg\ should roughly double the total number of blazars detected in \grays, and
will be enable multiwavelength observational campaigns to include the GeV region
of blazar spectra during the flight years of 2003-2006.  There is an additional
possibility that an indirect detection of the cosmic UV and optical photon
background can be made through the detection of extinctions in high-redshift
blazars above $\sim 20$ GeV.  \amsg\ will also likely observe GeV \gray\ emission
from one or more gamma ray bursts during its operational lifetime.

\amsg\ will also search for both line and continuum emission of \grays\ from the
region of the Galactic Center created by the annihilation of dark matter neutralinos.
Although the sensitivity to line emission appears marginal, there is nevertheless 
a finite, though small, region of halo/MSSM phase space which allows a detection by AMS.
However, a much larger region of dark-matter halo/MSSM parameter space can be constrained in
a search for continuum \grays\ by \amsg.

Finally, we note that even higher sensitivities can be reached with the addition
of a high granularity calorimeter below the magnet (option (b)); this will be the
subject of future work.

\end{document}